\documentclass{article}
\usepackage{spconf,amsmath,graphicx}
\usepackage{amsmath,graphicx}
\usepackage{booktabs}
\usepackage{caption}
\usepackage{subcaption}
\usepackage{dsfont}
\usepackage{bm}
\usepackage{hyperref}
\usepackage{setspace}
\usepackage{xcolor}

\title{An Experimental Study on Private Aggregation of Teacher Ensemble Learning for End-to-End Speech Recognition}
\name{%
\begin{tabular}{@{}c@{}}
Chao-Han Huck Yang$^{1, 2*}$\thanks{$^*$Work done at Georgia Tech and Amazon. Parts of this study were completed while the first author was an intern at Amazon.} \qquad
I-Fan Chen$^{2}$ \qquad Andreas Stolcke$^{2}$ \\ \qquad Sabato Marco Siniscalchi$^{1,3}$\qquad Chin-Hui Lee$^{1}$
\end{tabular}}
\address{$^1$Georgia Institute of Technology, USA and $^2$Amazon Alexa AI, USA \\ $^{3}$Department of Electronic Systems, NTNU, Trondheim, Norway}
\begin{document}
\maketitle
\begin{abstract}
Differential privacy (DP) is one data protection avenue to safeguard user information used for training deep models by imposing noisy distortion on privacy data. Such a noise perturbation often results in a severe performance degradation in automatic speech recognition (ASR) in order to meet a privacy budget $\varepsilon$. Private aggregation of teacher ensemble (PATE) utilizes ensemble probabilities to improve ASR accuracy when dealing with the noise effects controlled by small values of $\varepsilon$. We extend PATE learning to work with dynamic patterns, namely speech utterances, and perform a first experimental demonstration that it prevents acoustic data leakage in ASR training.
We evaluate three end-to-end deep models, including LAS, hybrid CTC/attention, and RNN transducer, on the open-source LibriSpeech and TIMIT corpora. PATE learning-enhanced ASR models outperform the benchmark DP-SGD mechanisms, especially under strict DP budgets, giving relative word error rate reductions between 26.2\% and 27.5\% for an RNN transducer model evaluated with LibriSpeech. We also introduce a DP-preserving ASR solution for pretraining on public speech corpora. 

\end{abstract}
\section{Introduction}
Automatic speech recognition (ASR)~\cite{ rabiner2007introduction} is  widely used in spoken language processing applications, such as smart device control~\cite{mcgraw2016personalized}, intelligent human-machine dialogue~\cite{stolcke2000dialogue}, and spoken language understanding \cite{baker2009developments}. To build ASR systems, a large collection of user speech~\cite{panayotov2015librispeech} is often required for training high-performance acoustic models. Protecting information privacy and measuring numerical privacy loss, such as whether data from a specific user is used for model training, are becoming critical and prominent research topics for on-device speech applications \cite{cheng2022personal, yang2021pate, cui2021federated, tomashenko2022voiceprivacy, yang2022mitigating, wu2021improving, lin2022speech}.

Differential privacy (DP) introduces a noise addition scheme for information protection characterized by \textbf{measurable privacy budgets}. The noise level is defined by a privacy budget (e.g., controlled by a minimum $\varepsilon$ value) in $\varepsilon$-DP~\cite{dwork2008differential, mironov2017renyi} for a given data set. Machine learning frameworks based on $\varepsilon$-DP have been shown effective against leakage of training data (e.g., human faces), model inversion attacks~\cite{fredrikson2015model} (MIA) and membership inference, in which a query-free algorithm is used to generate highly confident test data similar to its training set. Deploying $\varepsilon$-DP in speech applications based on deep models has recently attracted much interest~\cite{abadi2016deep}.
However, directly applying $\varepsilon$-DP perturbation on the training data could lead to severe performance (e.g., prediction accuracy) degradation~\cite{rajkumar2012differentially}. Therefore, the noise-enabled protection process needs to be designed carefully for incorporation into machine learning for training speech models. Most published DP-based approaches are still limited to recognition of isolated spoken commands \cite{yang2021pate}. Designing a continuous speech recognition system with $\varepsilon$-DP protection needs further investigation under a variety of privacy settings.  

In this work we use $\varepsilon$-DP protection to show that acoustic features used in ASR training data can been protected against model inversion attack (MIA)~\cite{fredrikson2015model}. We also demonstrate that $\varepsilon$-DP can prevent data leakage from a pretrained ASR model. %

\begin{figure*}[t!]
    \centering
    \includegraphics[width=0.85\linewidth]{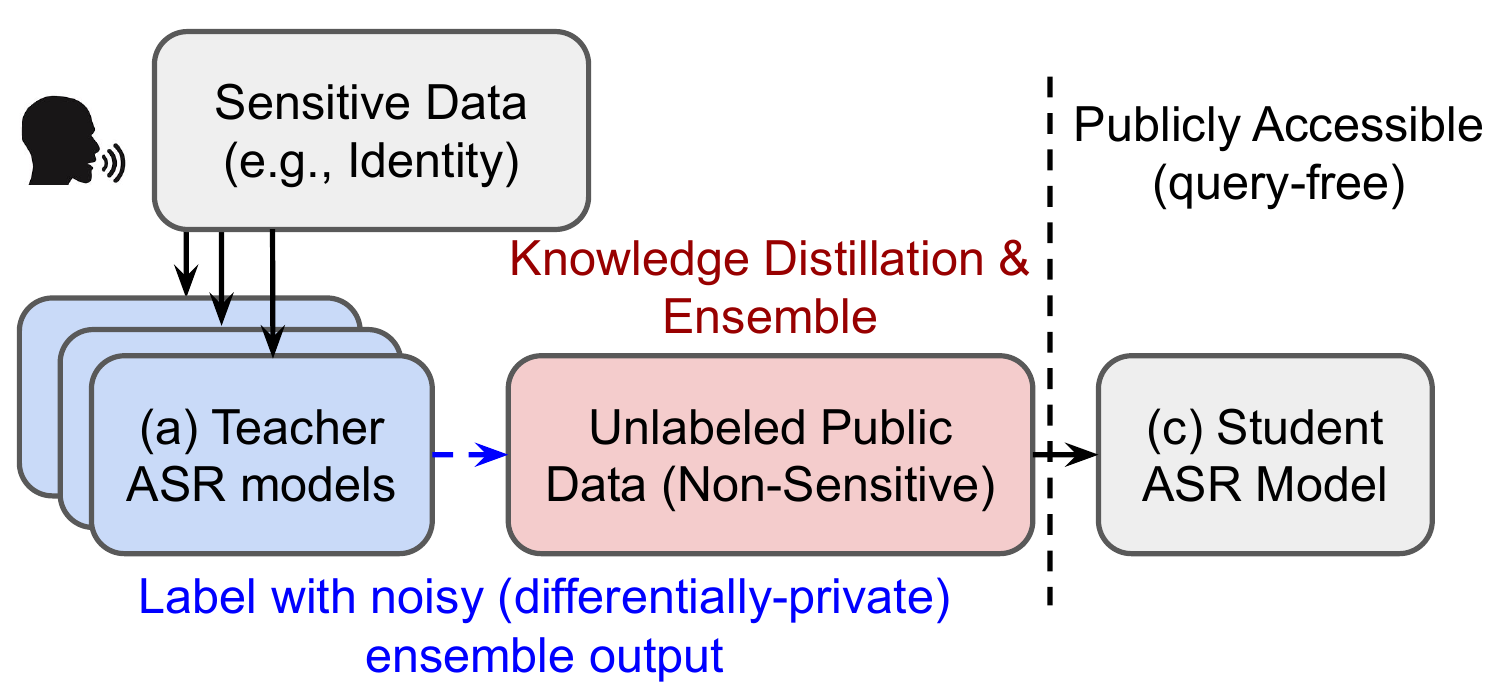}
    \caption{Proposed framework for utilizing private aggregation teacher ensemble (PATE) to train end-to-end ASR with $\varepsilon$-differential privacy. Teacher ASR models could also be combined with pretraining on public data (our second case study).}
    \label{fig:1}
\end{figure*}

Private aggregation of teacher ensembles (PATE)~\cite{papernot2016semi} learning is a recently proposed framework that aims to avoid accuracy loss in large-scale visual prediction models while guaranteeing $\varepsilon$-DP protection. PATE is guided by a teacher-student learning process, where multiple teachers make up an ensemble for knowledge transfer. The idea underlying PATE is to benefit from aggregated noisy outputs of teacher models to compartmentalize nonsensitive public data with $\varepsilon$-DP protections. PATE and related approaches~\cite{papernot2018scalable} were proven effective in mitigating model accuracy deterioration by employing an ensemble of teacher models with independent noise perturbations. Nonetheless, for ASR systems dealing with continuous speech, label stream prediction (e.g., by sequence level modeling) is required. In this study, we apply the PATE framework to continuous speech recognition and design different ensemble strategies to ensure $\varepsilon$-DP for end-to-end ASR models, including RNN transducer~\cite{graves2012sequence}, hybrid CTC/attention~\cite{kim2017joint}, and LAS (Listener, Attender, Speller)~\cite{chan2016listen} networks, as shown in the blue squares in Figure~\ref{fig:1}. Under a strict DP budget ($\varepsilon$=1), the PATE-trained end-to-end models maintain data privacy at the expense of only a slight increase in word error rate (WER). It also shows competitive advantages when compared to differentially private stochastic gradient descent (DP-SGD)~\cite{abadi2016deep, dwork2010boosting} benchmarks.

\section{Related Work}
Recent research efforts to preserve data privacy in an ASR system can be categorized into two groups: (i) systemic, such as federated learning \cite{leroy2019federated}, data isolation \cite{yang2021decentralizing}, and data encryption \cite{glackin2017privacy}, and (ii) algorithmic, mainly differentially private machine learning \cite{abadi2016deep}. In this section, we first summarize some of the privacy-preserving solutions proposed for ASR. Next, we describe the substratum of differential privacy devised for machine learning applications and discuss their difference to our proposed approach, while highlighting its key contributions.

\subsection{Privacy-Preserving Automatic Speech Recognition}
Federated machine learning algorithms~\cite{leroy2019federated,  dimitriadis2020federated, yang2021decentralizing} have been investigated in the ASR community to improve information protection. For instance, the average gradient method~\cite{dimitriadis2020federated} was deployed to update the model in ASR training. Vertical federated learning methods~\cite{yang2021decentralizing} show some other benefits from isolated features extractors under a heterogeneous computation. However, these system-level frameworks often rely on assumptions about the constrained accessibility of the malicious threats and barely provide universal and statistical justification for privacy guarantees.  Cryptographic encryption~\cite{glackin2017privacy, brasser2018voiceguard} and computation protocols~\cite{pathak2012privacy} are established techniques for privacy-preserving speaker identification. Meanwhile, these encryption algorithms and protocols do not consider privacy protection for training samples, which is a crucial element in developing machine learning models on a large scale. Lately, differentially-private stochastic gradient descent (DP-SGD)~\cite{abadi2016deep, dwork2010boosting} was introduced to allow quantitative measurements and further protect identity inference (e.g., of accent or speaking condition) by introducing an additive distortion during model training. In the remainder of the paper, a mathematical formulation of differential privacy is provided, as well as a study of its effectiveness for ASR.

\subsection{Differential Privacy Fundamentals}
The differential privacy mechanism~\cite{dwork2008differential} is a standard method to deploy algorithms with a target privacy guarantee.

\textbf{Definition 1.} A randomized algorithm $\mathcal{M}$ with domain $\mathcal{D}$ and range $\mathcal{R}$ is $(\varepsilon, \delta)$-differentially private if for any two neighboring inputs (e.g., speech data) $d, d^{\prime} \in \mathcal{D}$ and for any subset of output predictions (e.g., labels) $S \subseteq \mathcal{R}$, the following holds:
\begin{equation}
\operatorname{Pr}[\mathcal{M}(d) \in S] \leq e^{\varepsilon} \operatorname{Pr}\left[\mathcal{M}\left(d^{\prime}\right) \in S\right]+\delta.
\label{eq:1}
\end{equation}
The definition above produces a notion of privacy that can be expressed as a measure of the probabilistic difference of a specific outcome by a multiplicative factor, $e^{\varepsilon}$, and an additive amount, $\delta$. The $DP$ mechanism with post-processing~\cite{abadi2016deep} (e.g., batch-wise training) is under a general Renyi-divergence~\cite{mironov2017renyi} measurement with order $\alpha \in(1, \infty)$, called $\text{RDP}_{\alpha}$, for all neighboring data sets $d, d^{\prime}$:
\begin{equation}
\begin{array}{l}
 \text{RDP}_{\alpha}\left(\mathcal{M}(d) \| \mathcal{M}\left(d^{\prime}\right)\right) \\
=\frac{1}{\alpha-1} \log S_{\theta \sim \mathcal{M}\left(d^{\prime}\right)}\left[\left(\frac{p_{\mathcal{M}(d)}(\theta)}{p_{\mathcal{M}\left(d^{\prime}\right)}(\theta)}\right)^{\alpha}\right] \leq \varepsilon
\end{array}
\end{equation}
As $\alpha \rightarrow \infty$, $\text{RDP}_{\alpha}$ converges to the standard $(\varepsilon, 0)$-DP.
Both $\varepsilon$ and
$\delta$ should be non-negative. Considering $\delta\rightarrow$ 0 with only minor relaxation, a smaller value of $\varepsilon$ indicates a stronger $(\varepsilon, 0)$-differentially private guarantee. In other words, nearly equal probabilities in Eq.~(\ref{eq:1}) would be given from the neighboring inputs $d$ and $d^{\prime}$, which makes data identity much hard to infer. Moreover, learning from post-processing features (e.g., Mel-spectrum \cite{davis1980comparison, bridle1974experimental}) based on the 
speech data could also be differentially private, which has been shown by the theorem given in~\cite{dwork2008differential, abadi2016deep}.

\section{PATE Learning for ASR}
\label{sec:3}

\subsection{Noisy Teacher Ensembles for Acoustic Modeling}

We now describe PATE~\cite{papernot2016semi, papernot2018scalable} based acoustic modeling to enable $\varepsilon$-DP. First, an ensemble of teacher models is built by partitioning the training dataset into $I$ disjoint subsets: $\mathcal{D}_1 ,..., \mathcal{D}_I$. Next, these are used to train $I$ teacher models independently: $\mathcal{T}_1,..., \mathcal{T}_I$. Each such model is employed to generate frame-level acoustic model scores, which are then aggregated by weighted average and used as a teacher for student model training \cite{chebotar2016distilling, gao2021distilling, yang2021multi}.
For each frame of audio $x$, the final aggregated teacher model produces a vector of posteriors $\mathcal{T}_{\mathrm{ens}}(s \mid x)$ over context-dependent states $s$ computed as follows:
\begin{equation}
\mathcal{T}_{\mathrm{ens}}(s \mid x)=\sum_{i=1}^{I} w_{i} \mathcal{T}_{i}(s \mid x),
\label{eq:2}
\end{equation}
where $\mathcal{T}_{i}(s \mid x)$ is the posterior from the $i$th model, and $w_{i}$ is its weighting coefficient. The corresponding states from individual teacher models have the same dimension $J$ before the output alignment (e.g., considering a special silent character before alignment).

To ensure $\varepsilon$-DP under the PATE method, a random perturbation was introduced into the individual teachers' predictions ($\mathcal{T}_{i}$). We  revise Eq.~(\ref{eq:2}) to obtain a final ensemble from noisy teachers:
\begin{equation}
\mathcal{T}_{\operatorname{PATE}}(x,{\lambda})=\sum_{i=1}^{I} w_{i}\left(\mathcal{T}_{i}(x)+Y_{j}({\lambda})\right),
\label{eq:3}
\end{equation}
where $Y_{1}, ..., Y_{m}$ are i.i.d.\ Laplacian or Gaussian random variables with location $0$ and scale $\lambda^{-1}$. $\lambda$ refers to a privacy parameter that influences $(\epsilon, \delta)$-DP guarantees and for which a bound has been proven under composition theorems applicable to model aggregation~\cite{papernot2016semi, papernot2018scalable, gao2021distilling}.

As shown in Figure~\ref{fig:1}, the next PATE step is a process of knowledge transfer, where the noisy ensemble output is used to relabel a nonsensitive data set, with a total sample number $K$, which in turn is used to train a student model, $\mathcal{S}$. Both the prediction outputs and the trained student model's internal parameters are free from querying requests, which allows the only privacy cost to be associated with acquiring the training data for the student model. We evaluate two noisy aggregation processes, Gaussian NoisyMax (GNMax) and  Laplacian NoisyMax (LNMax) presented in previous studies~\cite{papernot2016semi, papernot2018scalable}. Under this setup, the student model is $(\varepsilon, 10^{-3})$-DP guaranteed using $\lambda$ from an analysis in~\cite{papernot2016semi, papernot2018scalable}. According to Eq.~(\ref{eq:3}), a large $\lambda$ refers to a \textbf{smaller $\varepsilon$, providing a stronger privacy guarantee}, but degrades the accuracy of the labels from the noisy maximum prediction output of the PATE function.

\subsection{Sequence-level Teacher Distillations}
To transfer the label data for knowledge distillation (KD) on a privacy-preserving student model, $\mathbf{h}_{\mathcal{T}}^{e}$ and $\mathbf{h}_{\mathcal{S}}^{e}$ denote the hidden vector representations of the teacher and student encoders respectively, and $P_{\mathcal{T}}\left(v \mid \mathbf{h}_{\mathcal{T}}^{d}\right)$ and $P_{\mathcal{S}}\left(v \mid \mathbf{h}_{\mathcal{S}}^{d}\right)$ the posterior probabilities computed by teacher and student models, respectively, in regards to the label $v$. $\mathcal{L}_{K D}$ is defined at the sequence-level as:
\begin{equation}
\mathcal{L}_{K D}=-\sum_{S} \sum_{n=1}^{N} P_{\mathcal{T}}\left(\hat{\mathbf{y}}_{n} \mid \mathbf{h}_{\mathcal{T}}^{e}\right) \log p_{\mathcal{S}}\left(\hat{\mathbf{y}}_{n} \mid \mathbf{h}_{\mathcal{S}}^{e}\right)
\end{equation}
with $\hat{\mathbf{y}}_{n}$ being the $n$th hypothesis from the set of $N$-best hypothesis derived from beam-search (i.e., beam width $=N$ ) for the teacher.

\subsection{Three Evaluated Acoustic Models}
\label{sec:3:3}
Three deep ASR models are adopted to investigate the impact of PATE. We consider homogeneously distributed settings, where the same network architecture is used in both teacher ($\mathcal{T}$) and student ($\mathcal{S}$) models in teacher-student learning.

\textbf{RNN transducer (RNN-T)}~\cite{graves2012sequence} utilizes a joint network to combine current acoustic observations from an encoder network and predictions based on previously recognized tokens. The encoder network is an RNN that converts the input acoustic features into a hidden representation for each frame of input. The prediction network generates output from previous nonblank output labels. The joint network computes output token logits. The blank symbol is treated as a possible output to account for the length mismatch between input and output token sequences, as in CTC~\cite{graves2006connectionist} models. The loss function of RNN-T~\cite{graves2012sequence} is computed as the negative log posterior of the output label sequence given the input acoustic feature.

\textbf{Hybrid CTC/attention (CTC/Att)}~\cite{kim2017joint} combines a sequence-to-sequence  (seq2seq)  attention-based~\cite{bahdanau2016end} model with a CTC loss~\cite{graves2006connectionist}. The encoder processes an input sequence and creates a hidden latent representation of the same length as the target sequence. The training data set contains target utterances composed of predefined vocabulary tokens (characters, tokens, or words). The CTC loss is computed from the predictions obtained from the encoder with sequence modeling.

\textbf{LAS}~\cite{chan2016listen} consists of 3 components: (i) a \textbf{listener} (encoder), which is similar to a standard acoustic model, takes a time-frequency representation of the speech input and uses a group of neural network layers to map the input to higher-level features; (ii) an \textbf{attender}, which takes encoder features as input to learn an alignment between the input features and the predicted subword units, where each subword is typically a grapheme or word piece; and finally, (iii) \textbf{speller} (decoder), which takes the output of the attention module and generates a probability for each hypothesized word.

\section{Experiments}

\subsection{Experimental Setup}
We consider two privacy-preserving conditions for ASR modeling with $\varepsilon$-DP measurement and protection:
\begin{enumerate}
    \item Parts of the training data require $\varepsilon$-DP protection, and
    \item All training data requires $\varepsilon$-DP protection,
\end{enumerate}
using two open-source speech corpora for training and evaluation: LibriSpeech~\cite{panayotov2015librispeech} and TIMIT~\cite{garofolo1993darpa}.

In Condition 1 ($\mathcal{C}$1), we assume both public and nonpublic data are from the same domain (e.g., same recording process and acoustic condition), but part of the training speech is associated with information that is potentially sensitive and needs protection. We select LibriSpeech and split the ``train-clean960'' set into 800 hours of ``sensitive'' data for learning teachers models, and 160 hours (as the nonsensitive data shown in Fig.~\ref{fig:1}) for training a $\varepsilon$-DP preserving student models with open access to end-users. We use the Librispeech ``test-clean'' portion to report word error rates (WERs).

In Condition 2 ($\mathcal{C}$2), we assume that two data sets coming from different domains can be accessed during acoustic modeling. One data set (LibriSpeech in our case) is publicly available in its entirety without privacy concerns, while the other data set (TIMIT in our case) is private and needs to be protected. We select different ASR models with pretraining on 800 hours of LibriSpeech used in $\mathcal{C}$1, and adopt the models for fine-tuning on the TIMIT training set (55 hours) as teacher models. The $\varepsilon$-DP preserving student ASR model is trained with a knowledge distillations process using the same unlabeled data as $\mathcal{C}$1. We use the 5-hour TIMIT validation data to report WERs.

\subsection{Baseline and PATE Results}

\begin{figure}[ht!]
        \centering
        \begin{subfigure}[b]{0.213\textwidth}
            \centering
            \includegraphics[width=\textwidth]{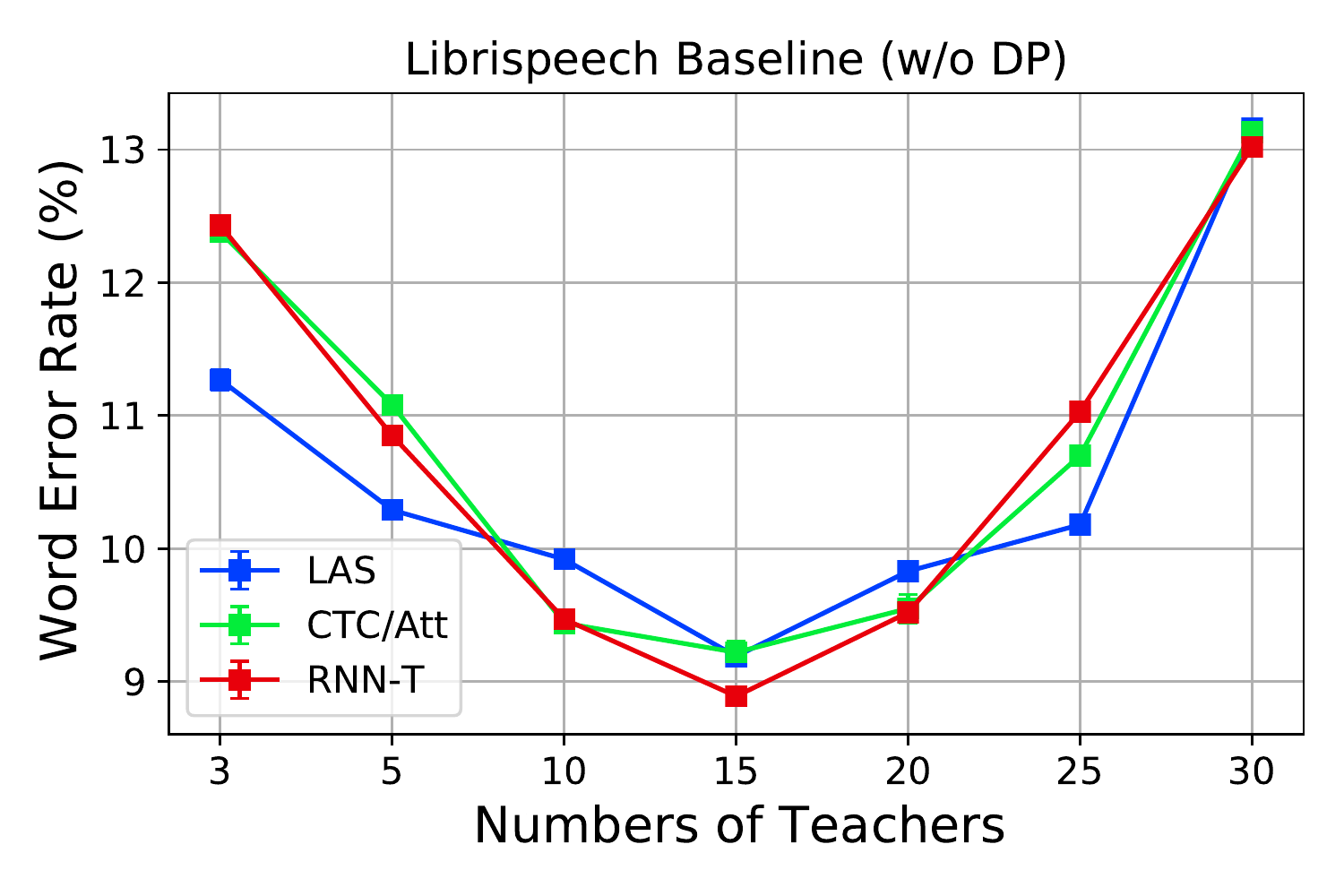}
            \caption{\small $\mathcal{C}$1: LibriSpeech}
        \end{subfigure}
        \quad~~~~
        \begin{subfigure}[b]{0.213\textwidth}
            \centering
            \includegraphics[width=\textwidth]{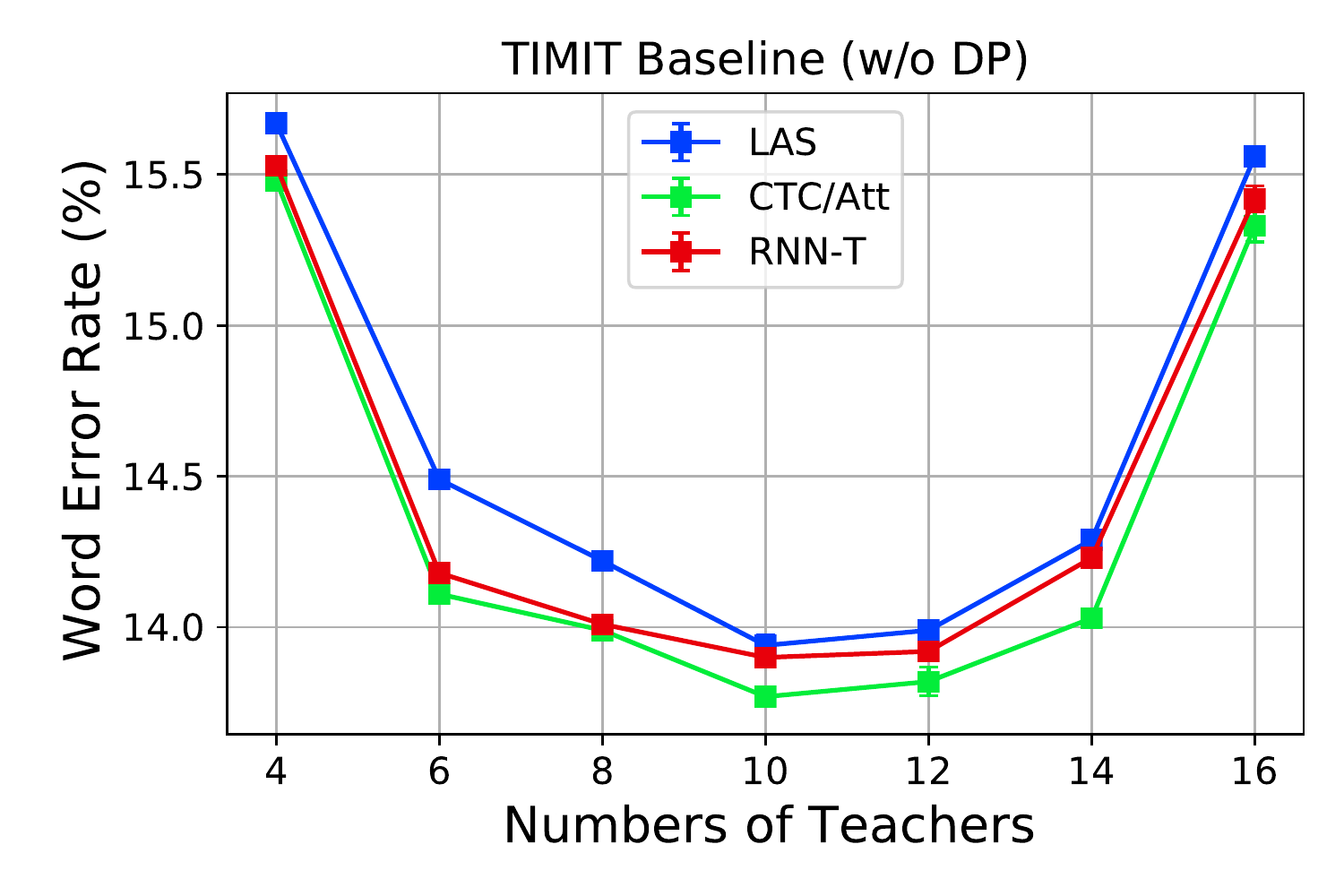}
            \caption{\small $\mathcal{C}$2: TIMIT}
        \end{subfigure}
        \caption{\small Baseline performance under ensemble training for ASR.}
        \label{fig:2:base}
    \end{figure}

\textbf{Ensemble baseline:} Before incorporating the aggregated ASR system for noisy teacher-student learning, we first evaluate how the size of the teacher ensemble influences general ASR performance. In previous work on PATE, researchers have demonstrated that having sufficient training data for each teacher model is crucial to ensure a high-performance end-to-end model. We start with the same number of teacher models reported in~\cite{papernot2016semi, papernot2018scalable} with $I$=20 in Eq.~(\ref{eq:2}) and evaluate with the various ASR models discussed in Section~\ref{sec:3:3}. The experimental results show that the data scale of the nonpublic training set used for teacher models could affect an optimal ensemble number. As shown in Fig.~\ref{fig:2:base}(a), the selected ASR models have lower WERs ($<10\%$) with $10$ to $20$ teachers in $\mathcal{C}$1 (training with LibriSpeech 800 hours as private). For $\mathcal{C}$2 (training with TIMIT 55 hours as private) shown in Fig.~\ref{fig:2:base}(b), the desired number of teachers becomes smaller ($10$ to $12$) to obtain competitive WER performance ($<14\%$). We fix the number of teachers at $15$ in $\mathcal{C}$1 and $10$ in $\mathcal{C}$2, for all following PATE-based experiments.

\begin{figure}[ht!]
        \centering
        \begin{subfigure}[b]{0.213\textwidth}
            \centering
            \includegraphics[width=\textwidth]{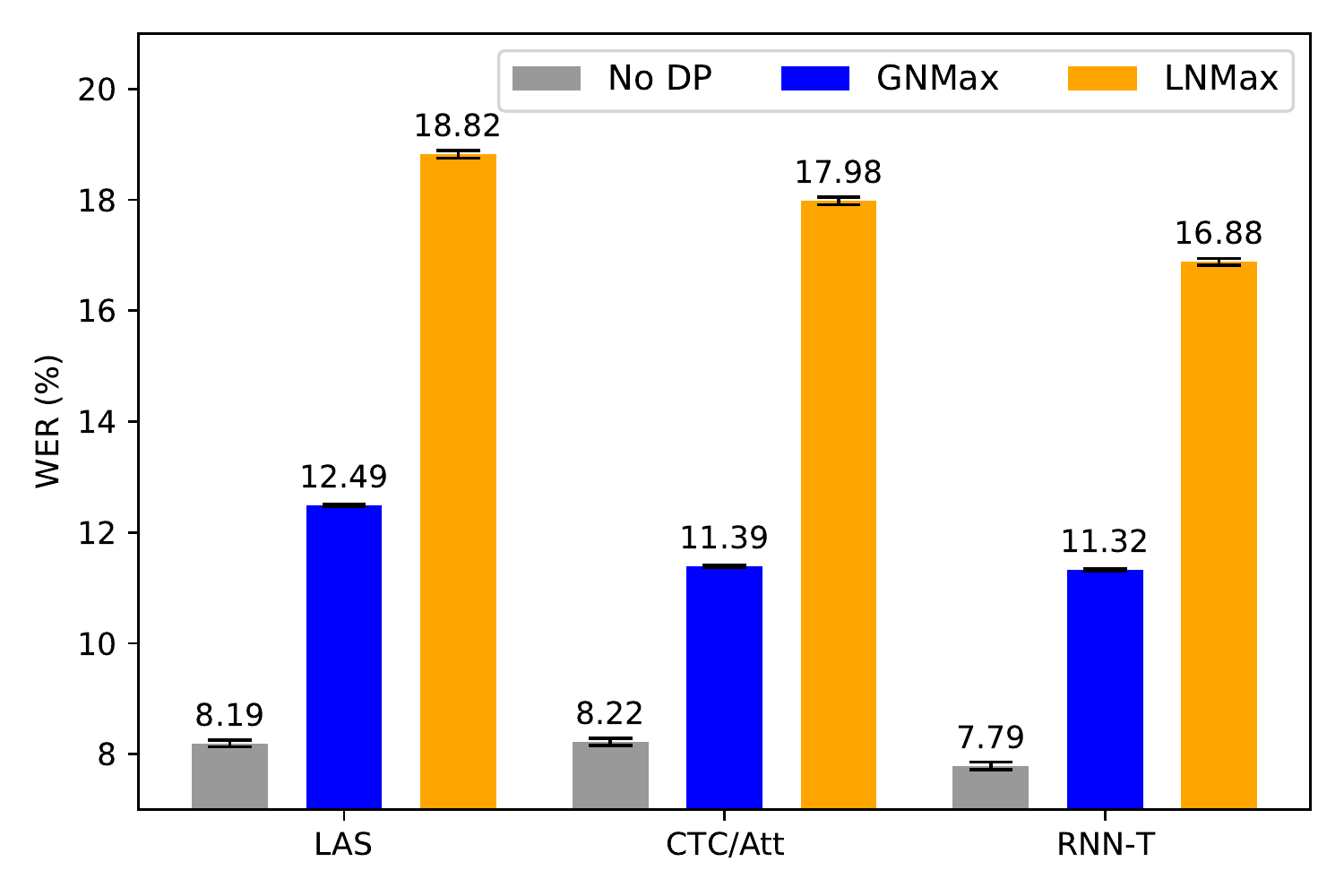}
            \caption{\small $\mathcal{C}$1: LibriSpeech}
        \end{subfigure}
        \quad~~~~
        \begin{subfigure}[b]{0.213\textwidth}
            \centering
            \includegraphics[width=\textwidth]{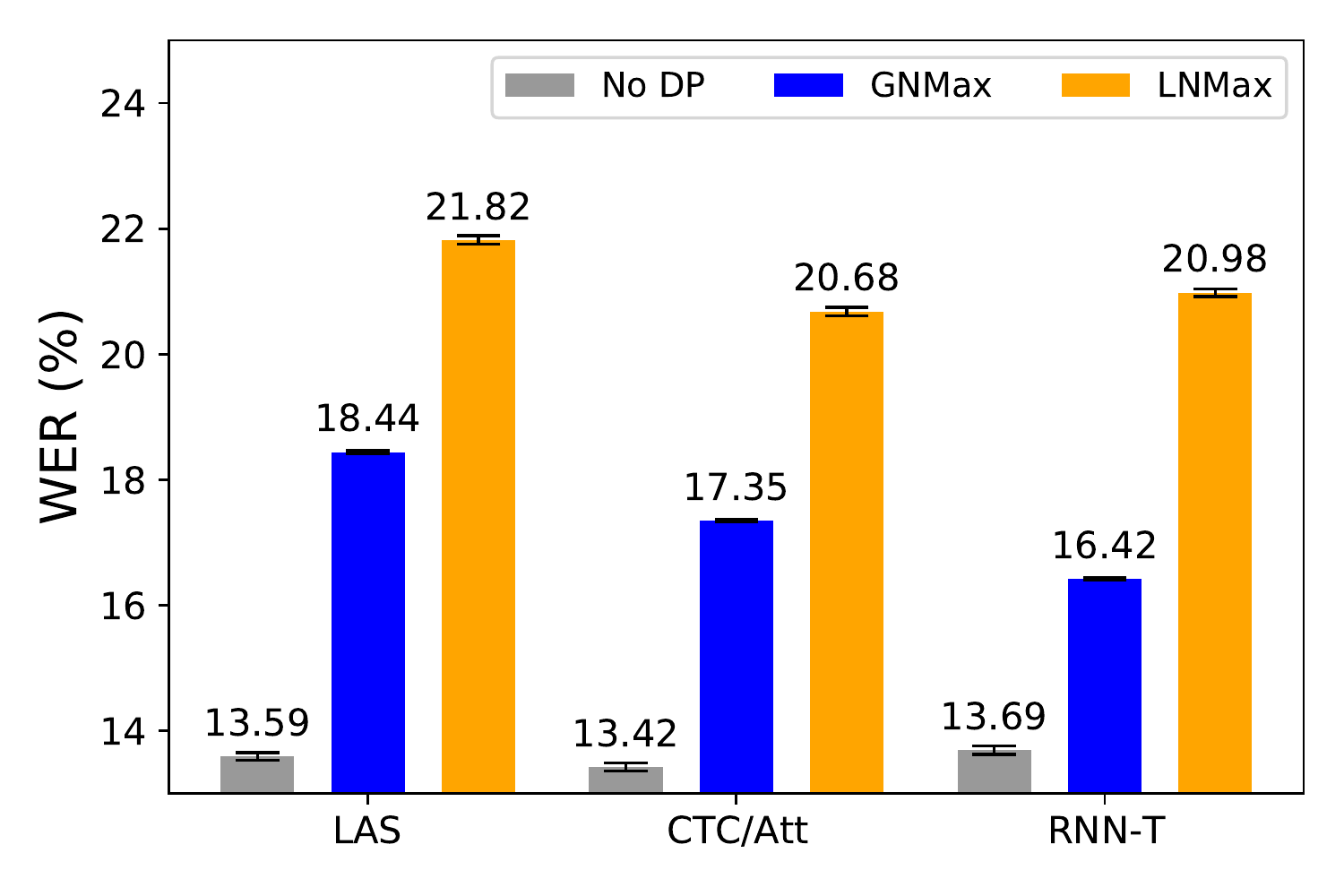}
            \caption{\small $\mathcal{C}$2: TIMIT}
            \label{fig:62}
        \end{subfigure}
        \caption{\small Student ASR model performance with two additive noisy mechanisms, GNMax and LNMax, under PATE for teacher training.}
        \label{fig:3:en}
    \end{figure}

\textbf{Additive noise selection:} to study how the $\varepsilon$-DP based noise injection could impact the ASR performances, we compare two additive noise mechanisms, GNMax (with Gaussian noise) and LNMax (with Laplacian noise), with an empirical DP budget of $\varepsilon$=10, as has been reported for different on-device applications~\cite{zhu2020federated}. We follow the PATE framework proposed in Section~\ref{sec:3} to obtain the final student ASR models with $\varepsilon$-DP protection. As shown in Fig.~\ref{fig:3:en}, both PATE-GNMax (\textcolor{blue}{blue}) and PATE-LNMax (\textcolor{orange}{orange}) show absolute WER degradation of $2.97\%$ to $8.22\%$ ($\mathcal{C}$1) and $2.73\%$ to $8.23\%$ ($\mathcal{C}$2) when compared to student model training with clean aggregated teacher models (\textcolor{gray}{gray}). RNN-T based ASR with GNMax achieves the best performance under the $\varepsilon$=10 privacy budget, where PATE-GNMax shows an average of $4.32\pm0.12$\% absolute WER reduction compared with PATE-LNMax results.

\begin{figure}[ht!]
        \centering
        \begin{subfigure}[b]{0.213\textwidth}
            \centering
            \includegraphics[width=\textwidth]{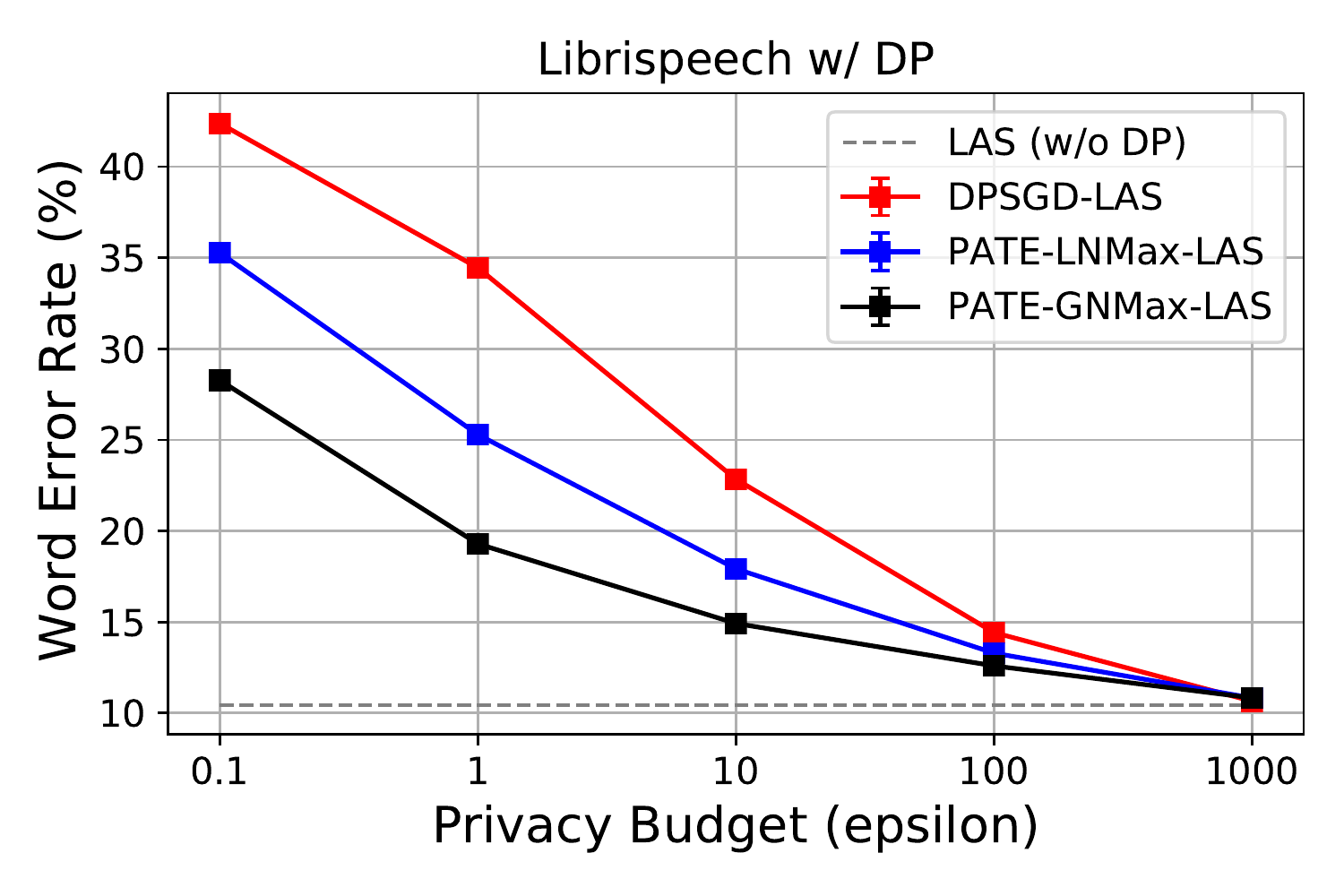}
            \caption{\small $\mathcal{C}$1: LibriSpeech}
        \end{subfigure}
        \quad~~~~
        \begin{subfigure}[b]{0.213\textwidth}
            \centering
            \includegraphics[width=\textwidth]{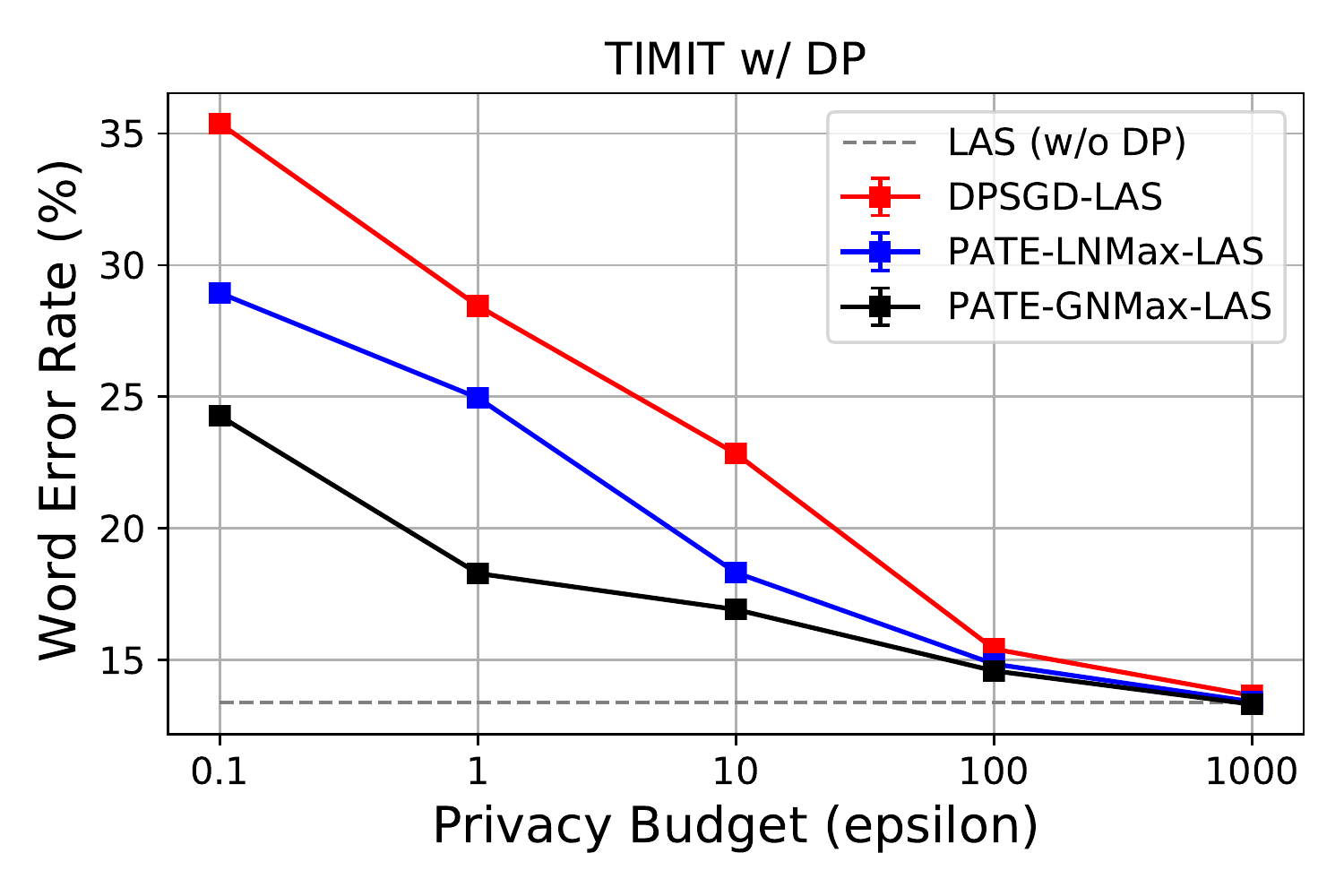}
            \caption{\small $\mathcal{C}$2: TIMIT}
        \end{subfigure}
        \caption{\small LAS results with PATEs and DP-SGD.}
        \label{fig:4}
    \end{figure}

\begin{figure}[ht!]
        \centering
        \begin{subfigure}[b]{0.213\textwidth}
            \centering
            \includegraphics[width=\textwidth]{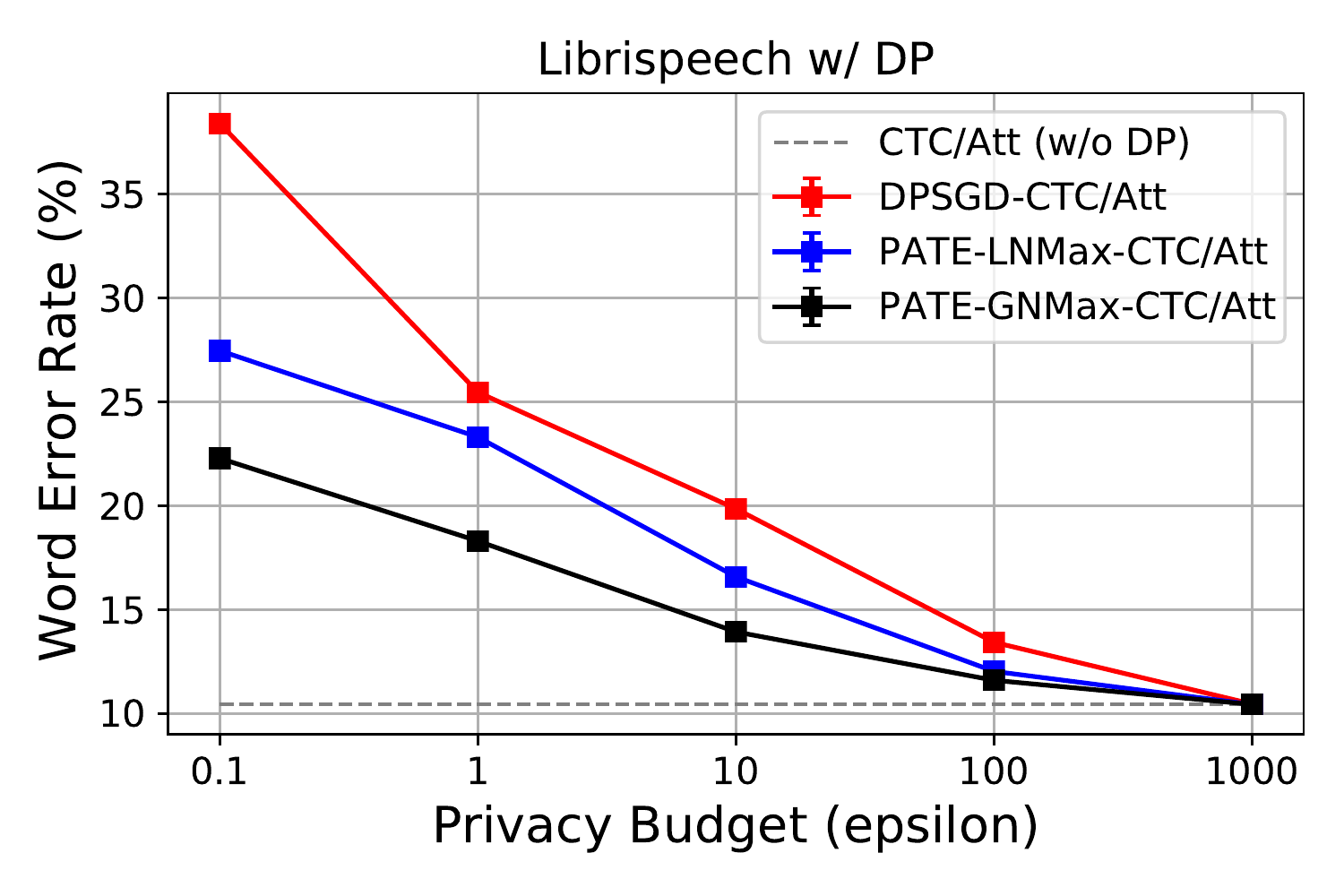}
            \caption{\small $\mathcal{C}$1: LibriSpeech}
        \end{subfigure}
        \quad~~~~
        \begin{subfigure}[b]{0.213\textwidth}
            \centering
            \includegraphics[width=\textwidth]{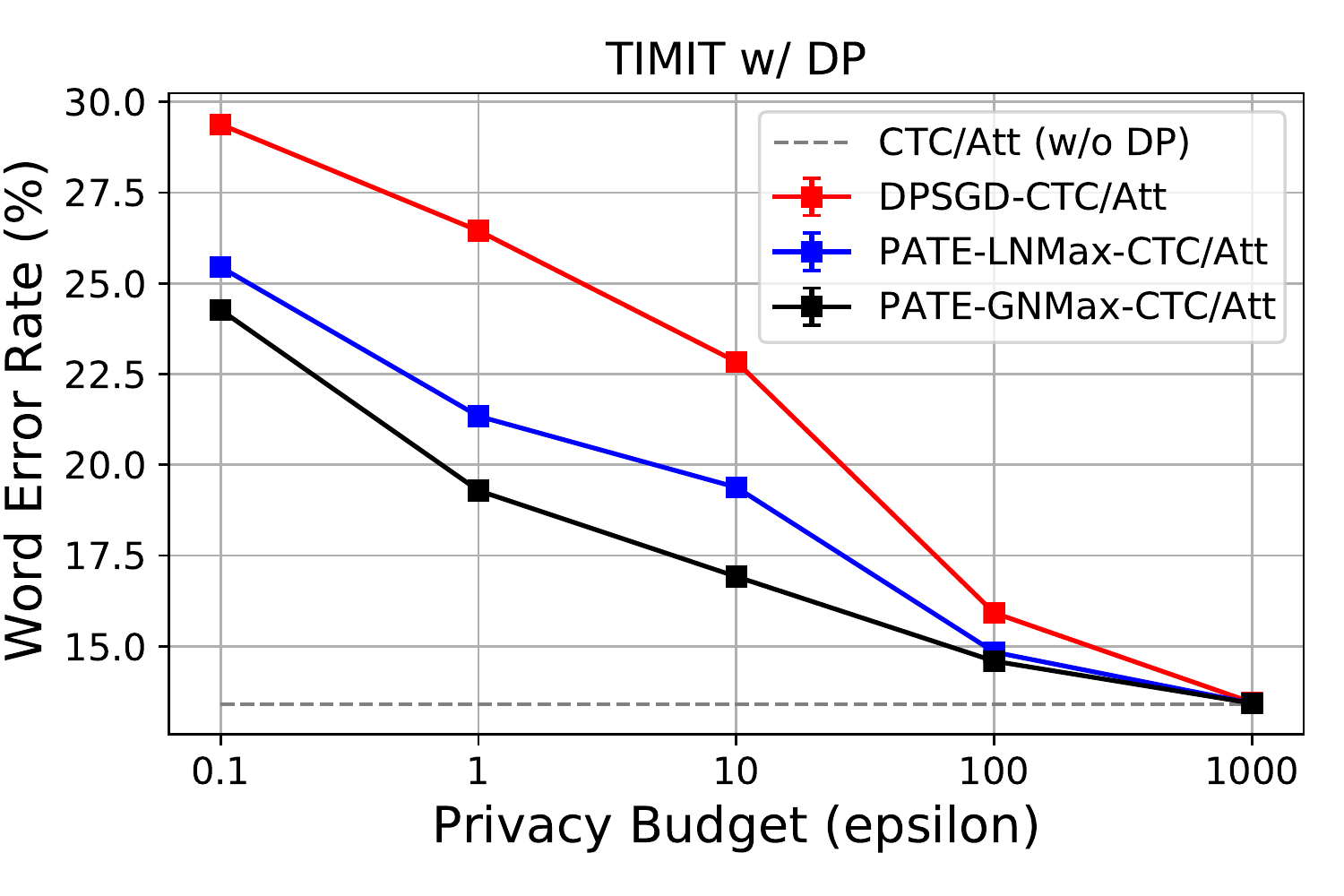}
            \caption{\small $\mathcal{C}$2: TIMIT}
        \end{subfigure}
        \caption{\small CTC/Attention results with PATEs and DP-SGD.}
        \label{fig:5}
    \end{figure}

\begin{figure}[ht!]
        \centering
        \begin{subfigure}[b]{0.213\textwidth}
            \centering
            \includegraphics[width=\textwidth]{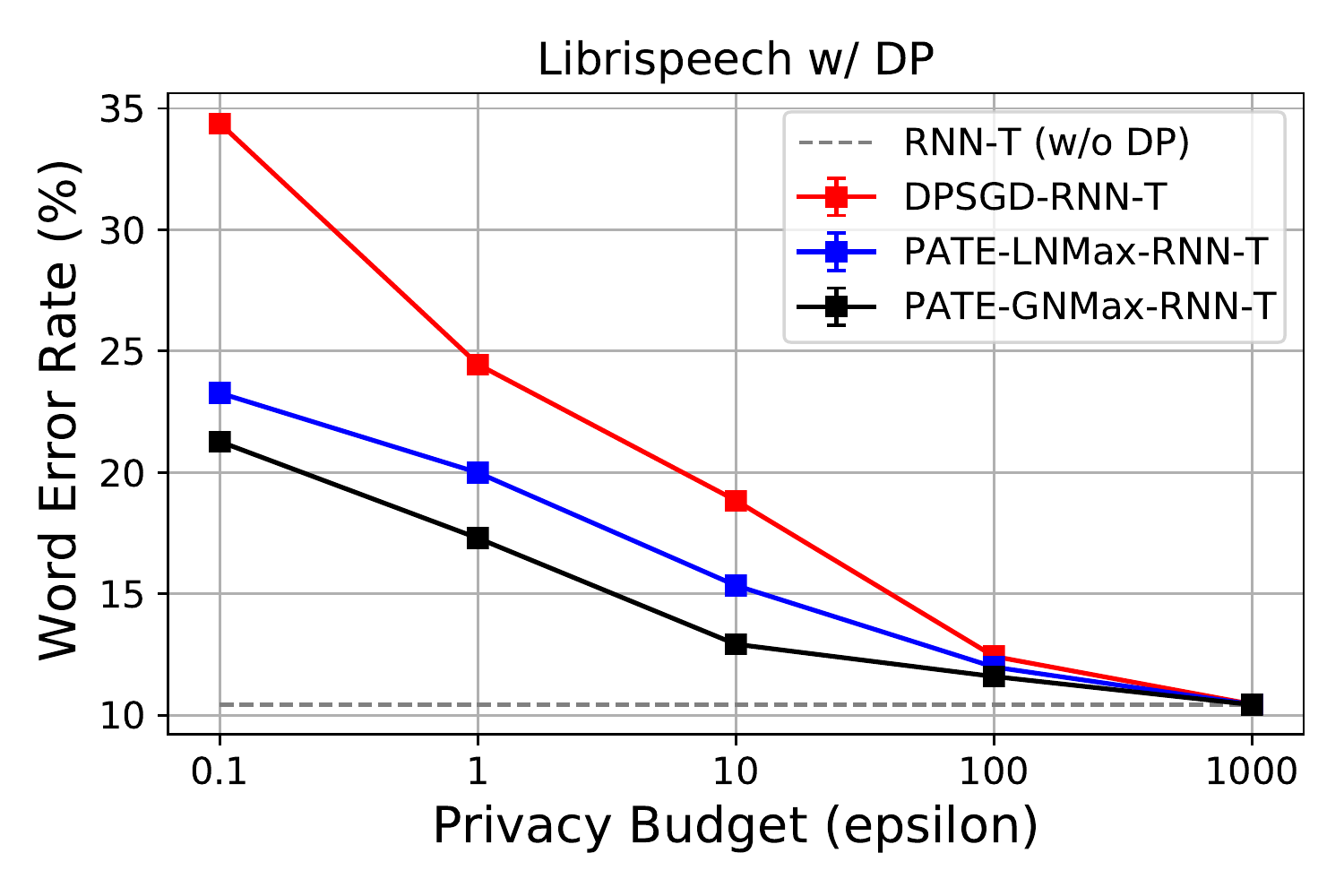}
            \caption{\small $\mathcal{C}$1: LibriSpeech}
        \end{subfigure}
        \quad~~~~
        \begin{subfigure}[b]{0.213\textwidth}
            \centering
            \includegraphics[width=\textwidth]{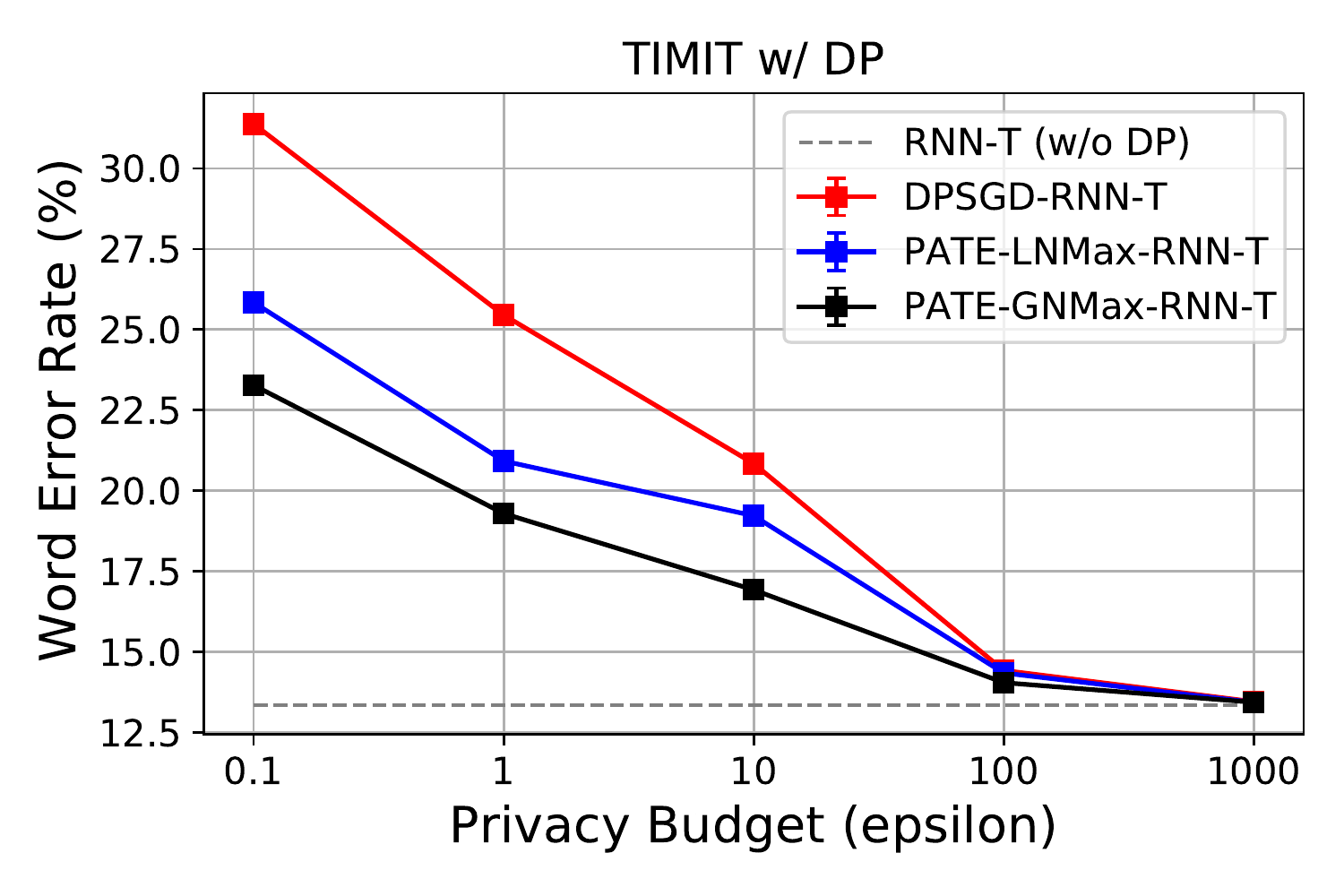}
            \caption{\small $\mathcal{C}$2: TIMIT}
        \end{subfigure}
        \caption{\small RNN-T results with PATEs and DP-SGD.}
        \label{fig:6}
    \end{figure}

\textbf{PATE for ASR with different privacy budgets:} We evaluate the proposed PATE-based algorithms at different noise levels, which represent different target privacy budgets  ($\lambda=\frac{K}{2\varepsilon}$). When compared with the benchmark $\varepsilon$-DP preserving DP-SGD~\cite{abadi2016deep} algorithm, the results in Figures~\ref{fig:4} (LAS)~\ref{fig:5} (Hybrid CTC/Att), and~\ref{fig:6} (RNN-T) demonstrate that PATE under the GNMax mechanism can achieve good performance when combined with different end-to-end architectures. They show $5.45\pm0.23\%$ and $9.81\pm0.43\%$ average WER reduction when comparing PATE-LNMax against DP-SGD~\cite{abadi2016deep} powered $\varepsilon$-DP preserving ASR models. They also highlight that the deployed models suffer major WER increases for privacy budgets $\varepsilon \leq 10$. The RNN-T based PATE model outperforms both LAS and Hybrid CTC/attention based ASR models when $\varepsilon\leq 100$. As an extreme case, the ASR model starts to converge to its clean ensemble when $\varepsilon \rightarrow 1000$ in both $\mathcal{C}$1 and $\mathcal{C}$2 conditions.

 \vspace{-2mm}

\subsection{A Case Study for $\varepsilon$-DP Protection Against MIA}

\begin{figure}[ht!]
        \centering
        \begin{subfigure}[b]{0.213\textwidth}
            \centering
            \includegraphics[width=\textwidth]{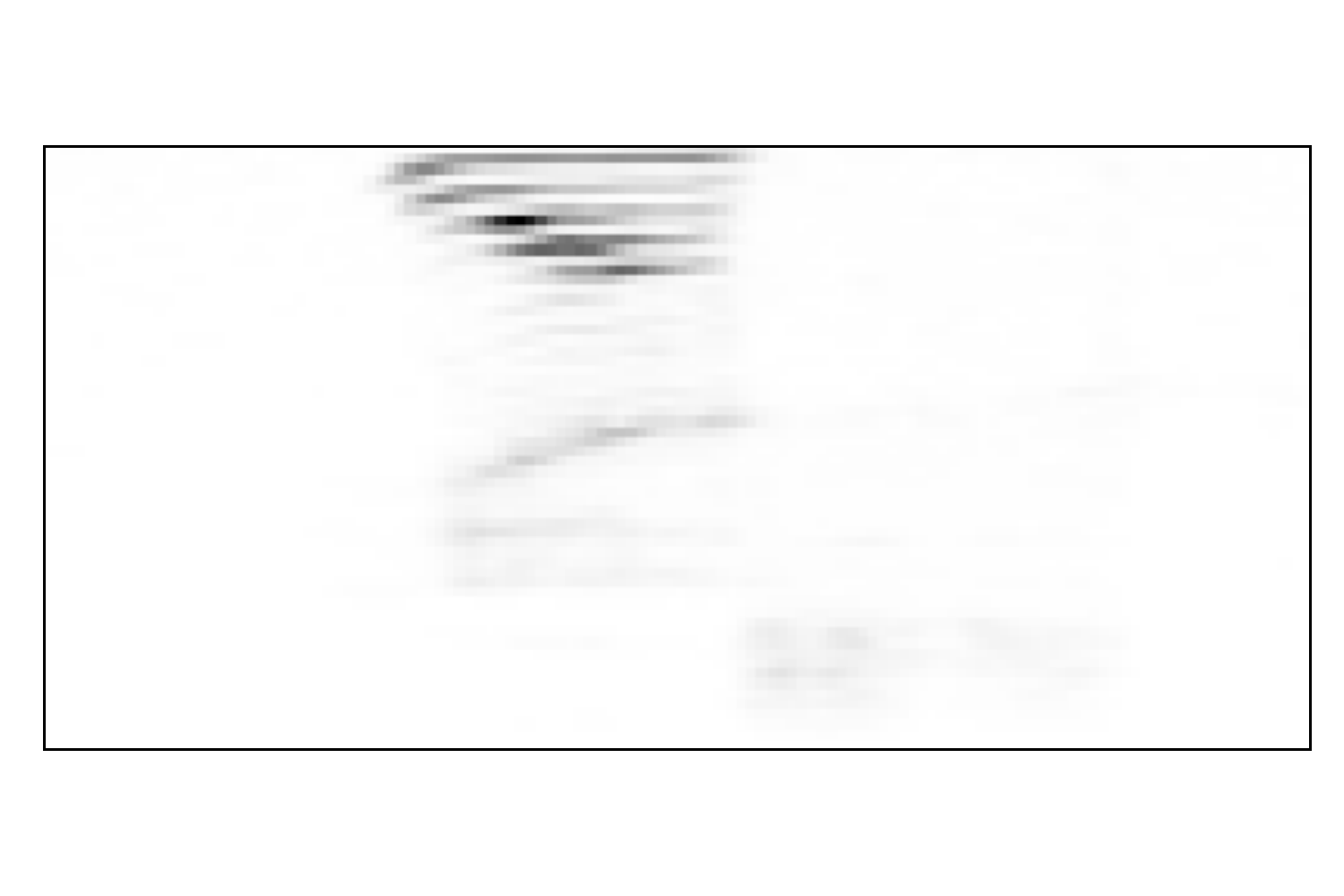}
            \caption{\small original (no MIA)}
            \label{fig:71}
        \end{subfigure}
        \quad~~~~
        \begin{subfigure}[b]{0.213\textwidth}
            \centering
            \includegraphics[width=\textwidth]{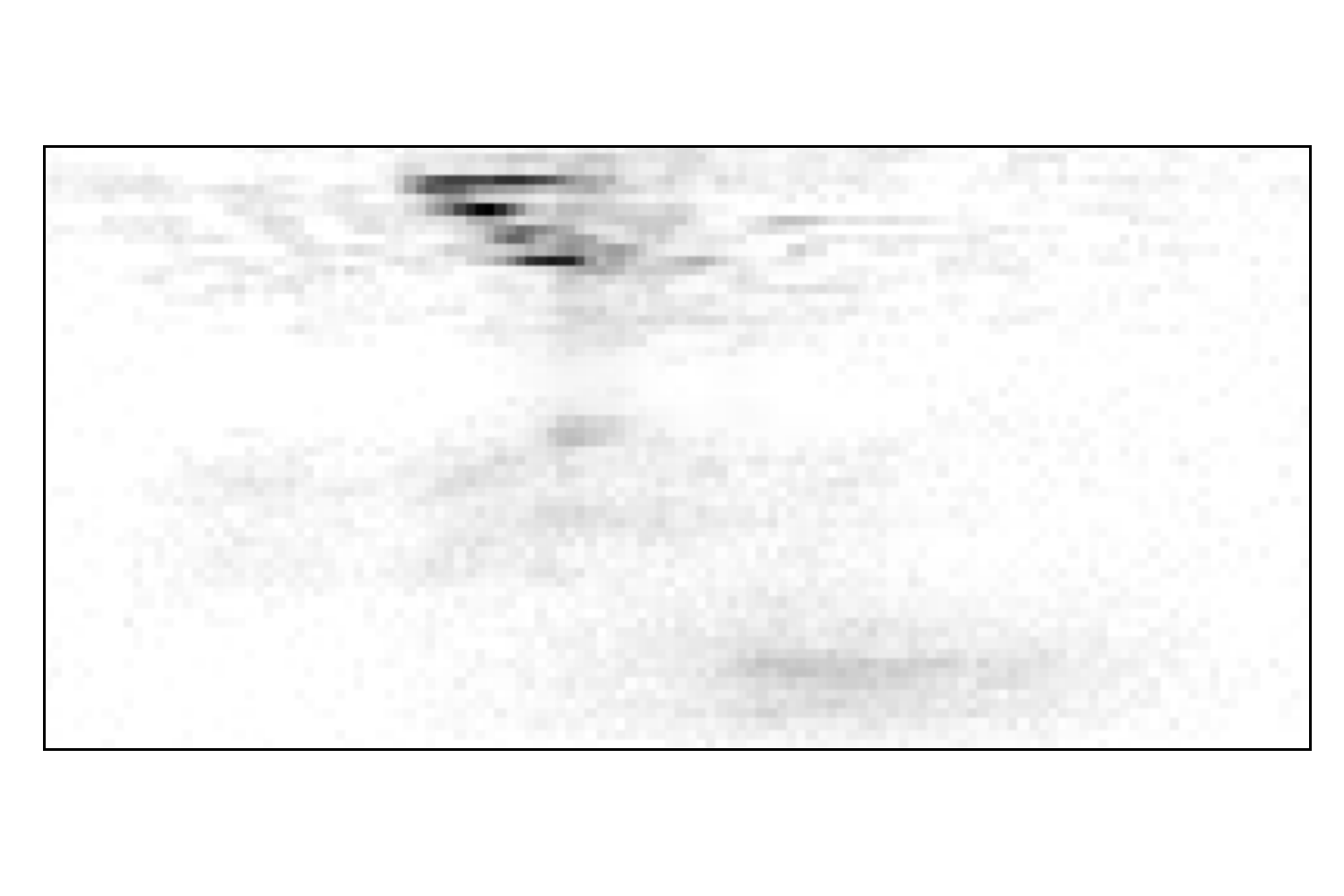}
            \caption{\small  MIA without DP ($\varepsilon$ = $\infty$) }
            \label{fig:72}
        \end{subfigure}
        \vskip\baselineskip
        \begin{subfigure}[b]{0.213\textwidth}
            \centering
            \includegraphics[width=\textwidth]{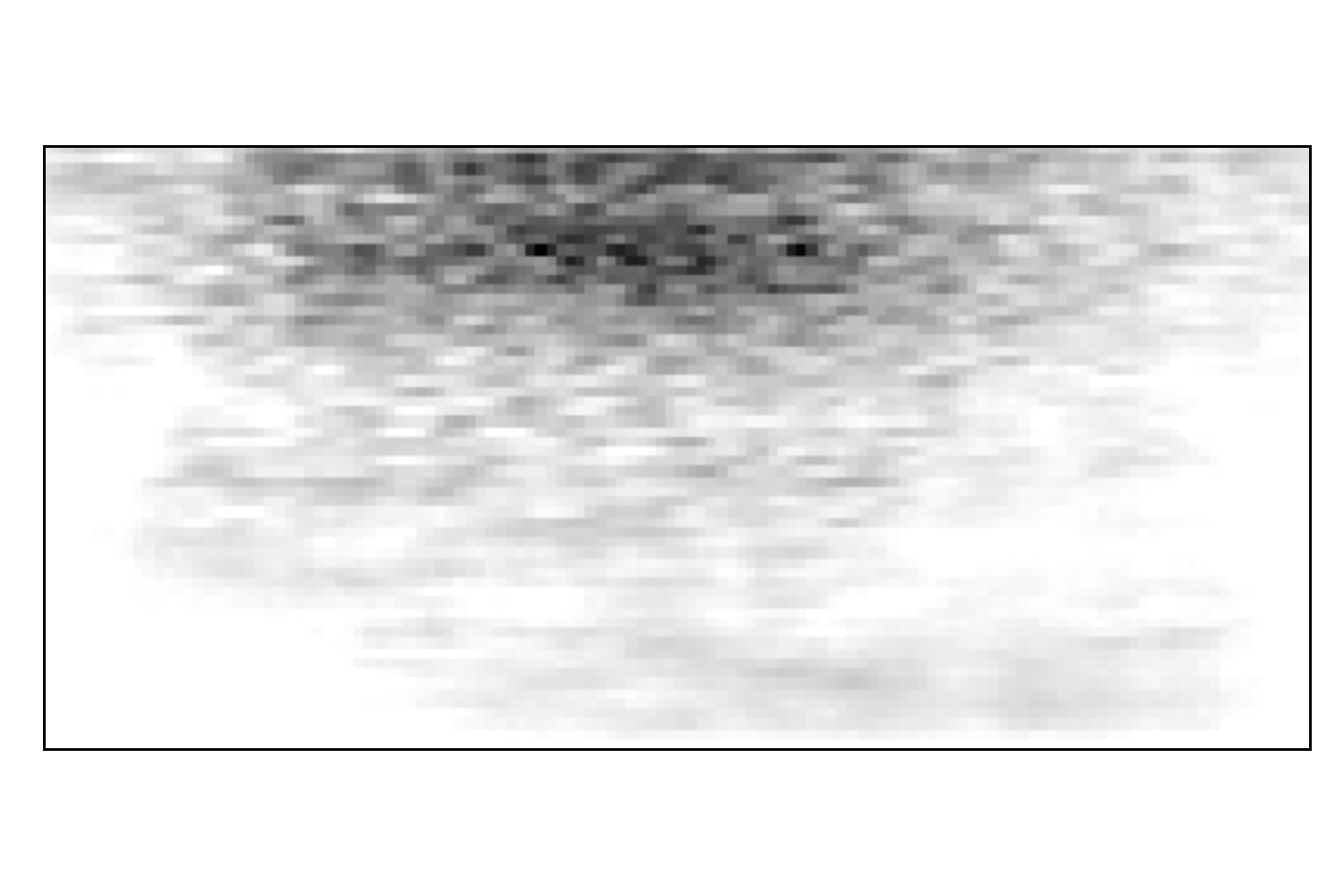}
            \caption{\small MIA with DP ($\varepsilon$ = 10)}
            \label{fig:73}
        \end{subfigure}
        \quad~~~~
        \begin{subfigure}[b]{0.213\textwidth}
            \centering
            \includegraphics[width=\textwidth]{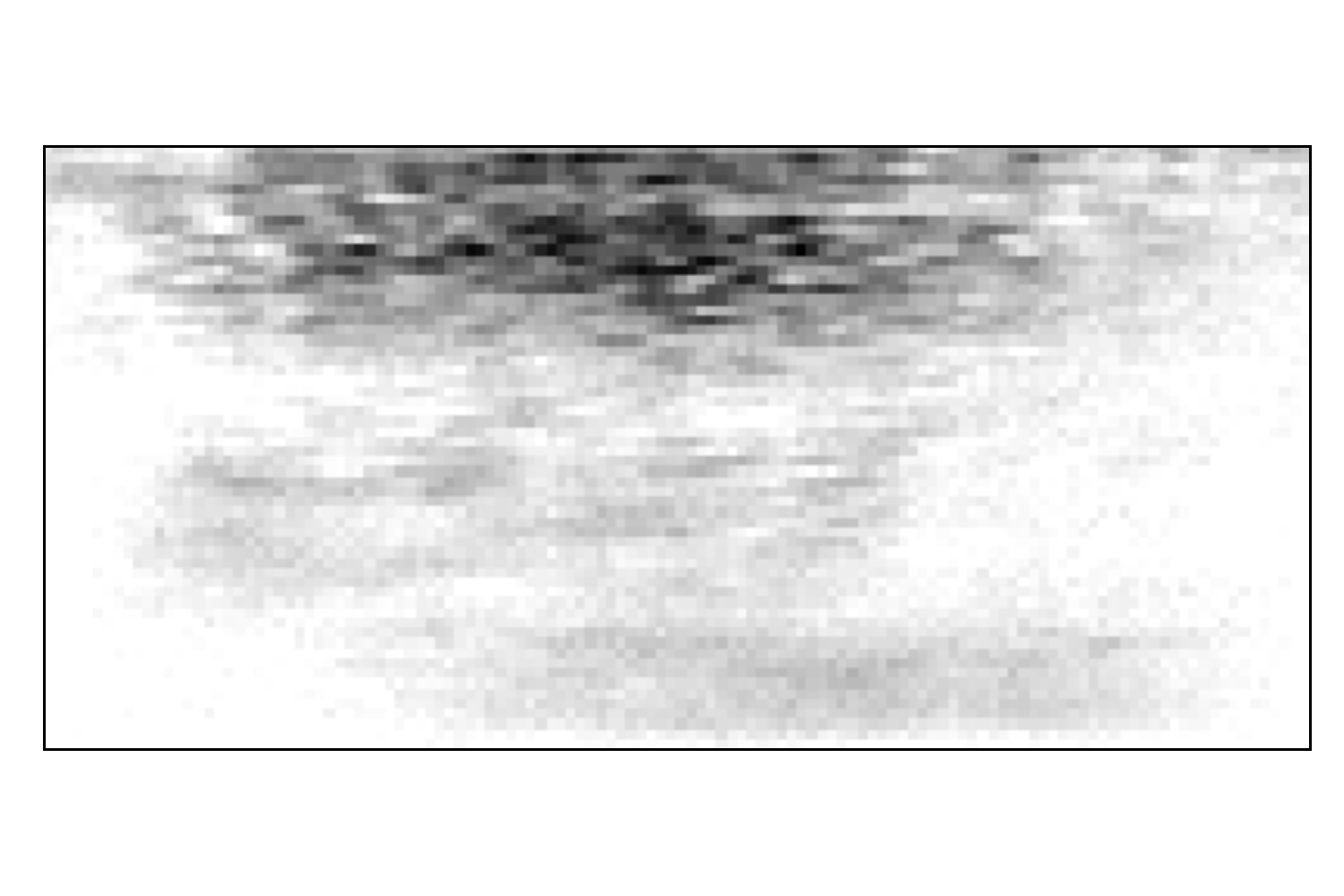}
            \caption{\small  MIA with DP ($\varepsilon$ = 1)}
            \label{fig:74}
        \end{subfigure}
        \caption{\small Model inversion attack (MIA) against ASR with a target word of ``stop.'' (c) and (d) show effective $\varepsilon$-DP protection.}
        \label{fig:7}
    \end{figure}

As a first step toward incorporating $\varepsilon$-DP into ASR acoustic modeling, we demonstrate how PATE can help protect user information in the training set, as used for training the teacher models. The demonstration considers the major privacy leakage method based on model inversion attack (MIA)~\cite{fredrikson2015model}, which utilizes the maximum a posteriori principle and constructs the input features that maximize the likelihood of observed model output queries (e.g., using an on-device API). We select the RNN-T based ensemble model trained under $\mathcal{C}$1 for a privacy-preserving case study against MIA with 10,000 service queries. We use the MIA algorithm to maximize the likelihood of a target output word of ``stop'' and observe its reconstructed high-confidence input audio.  As shown in Fig.~\ref{fig:7}, MIA can generate inverse Mel-spectrum output as in Fig.~\ref{fig:7}(b), similar to an original utterance clip shown in Fig.~\ref{fig:7}(a), presumably including speech characteristics unique to the speaker. However, $\varepsilon$-DP shows its effectiveness in preventing this information recovery attack with $\varepsilon \leq10$, as shown Fig.~\ref{fig:7}~(c) and Fig.~\ref{fig:7}~(d). This demonstrates how PATE-trained ASR models can successfully protect user information.

\vspace{-3mm}

\section{Conclusion}
We have conducted a first study in privacy-preserving ASR applying the PATE framework for $\varepsilon$-differential privacy to three popular ASR architectures. For all the DP budgets tested, PATE greatly outperforms previously proposed DP-SGD mechanisms as benchmarks, especially under strict DP budgets ($\varepsilon$=1). Using the best RNN-T model, we also demonstrated that the proposed $\varepsilon$-DP preserving PATE models can protect against potential private data leakage from the training set using
model inversion attacks. 
\clearpage
\bibliographystyle{IEEEbib}
\bibliography{refs}

\end{document}